\documentclass[aps,pre,twocolumn,superscriptaddress,showpacs,showkeys ]{revtex4-1}

\usepackage{graphicx}                   
\usepackage{hyperref}                   
\usepackage{amsmath,amssymb}            
\usepackage{epstopdf}
\usepackage{subcaption}					

\usepackage{color}
\definecolor{orange}{rgb}{1,0.5,0}
\definecolor{brown}{rgb}{0.65, 0.16, 0.16}
\definecolor{phlox}{rgb}{0.87, 0.0, 1.0}

\graphicspath{{figs/}}					

\begin{document}

\title{Gaussian Free Field in the background of correlated random clusters, formed by metallic nanoparticles}

\author{J. Cheraghalizadeh}
\affiliation{Department of Physics, University of Mohaghegh Ardabili, P.O. Box 179, Ardabil, Iran}
\email{jafarcheraghalizadeh@gmail.com}

\author{M. N. Najafi*}
\affiliation{Department of Physics, University of Mohaghegh Ardabili, P.O. Box 179, Ardabil, Iran}
\email{morteza.nattagh@gmail.com}

\author{H. Mohammadzadeh}
\affiliation{Department of Physics, University of Mohaghegh Ardabili, P.O. Box 179, Ardabil, Iran}
\email{h.mohammadzadeh@gmail.com}

\begin{abstract}
The effect of metallic nano-particles (MNPs) on the electrostatic potential of a disordered 2D dielectric media is considered. The disorder in the media is assumed to be white-noise Coulomb impurities with normal distribution. To realize the correlations between the MNPs we have used the Ising model with an artificial temperature $T$ that controls the number of MNPs as well as their correlations. In the $T\rightarrow 0$ limit, one retrieves the Gaussian free field (GFF), and in the finite temperature the problem is equivalent to a GFF in iso-potential islands. The problem is argued to be equivalent to a scale-invariant random surface with some critical exponents which vary with $T$ and correspondingly are correlation-dependent. Two type of observables have been considered: local and global quantities. We have observed that the MNPs soften the random potential and reduce its statistical fluctuations. This softening is observed in the local as well as the geometrical quantities. The correlation function of the electrostatic and its total variance are observed to be logarithmic just like the GFF, i.e. the roughness exponent remains zero for all temperatures, whereas the proportionality constants scale with $T-T_c$. The fractal dimension of iso-potential lines ($D_f$), the exponent of the distribution function of the gyration radius ($\tau_r$), and the loop lengths ($\tau_l$), and also the exponent of the loop Green function $x_l$ change in terms of $T-T_c$ in a power-law fashion, with some critical exponents reported in the text. Importantly we have observed that $D_f(T)-D_f(T_c)\sim\frac{1}{\sqrt{\xi(T)}}$, in which $\xi(T)$ is the spin correlation length in the Ising model.
\end{abstract}

\pacs{05., 05.20.-y, 05.10.Ln, 05.45.Df}
\keywords{Gaussian free field, metallic random clusters, Ising correlations, percolation lattice}

\maketitle

\section{Introduction}

The effect of environmental disorder is a long-standing problem in the condensed matter systems. The disorder, produced by doping neutral metallic nanoparticles (MNPs) in a $d$-dimensional ($d=$ two or three) solid state system, has an impressive impact in the transport of the media and can be tracked in both quantum and classical levels. The examples are incorporation of MNPs in graphene~\cite{moisala2003role,subrahmanyam2010study,zhou2009situ,kamat2009graphene}, enhancement of optical properties in optical devices using MNPs~\cite{miller2013nonlinear,haglund1993picosecond,shalaev1998nonlinear,cai2005superlens,kravets2010plasmonic,kim2017ultrafast,kim2018dielectric,kim2013nondegenerate} and the random laser~\cite{meng2013metal}, improving photocatalytic activity of TiO$_2$ films~\cite{subramanian2001semiconductor}, its application in the sensor technology~\cite{franke2006metal} and solar cells~\cite{huang2013mitigation,yu2017effects} and other thin films~\cite{li2016electrostatically}. MNPs cause a wide variety of physical phenomena in the condensed matter systems, such as the Anderson localization~\cite{arya1986anderson}, catalytic materials when incorporated to the porous media~\cite{white2009supported}, the effect of MPNs on organic nano-floating-gate memory devices~\cite{kang2013printed}, and on the acoustic vibrations in dielectric shells~\cite{mongin2011acoustic}. The subject has been vastly investigated in the literature~\cite{hui1986complex,perdew1993formation}. In many cases, in addition to the doped neutral MNPs, Coulomb disorders (impurities) are also present which make the problem theoretically challenging. Commonly the time scale of the dynamics of these impurities is very larger than the typical electronic time scales, which makes it permissible to take them into account as quenched. The other complexity of such systems is the formation pattern of the metallic clusters. It is well-known that the MNPs are not independent of each other, but their correlations cause them to aggregate and form some connected metallic clusters, i.e. one deals with metallic islands in the dielectric media. Despite of the huge theoretical and experimental literature concerning the subject, there is very limited knowledge on the effect of MNPs on the properties of the media. An overall agreement is also absent on the formation pattern of these clusters, since the correlation between the MNPs are not known which are controlled by their cohesive energies and also the media around. The Poisson equation for the dielectric media with uncorrelated Coulomb disorders in the absence of the MNP clusters in two-dimensions is mapped to the Gaussian free field (GFF) problem and the Coulomb gas in general.\\
The most general Coulomb gas is a gas of particles that carry a scalar quantized electric and magnetic charges with no external electric or magnetic fields. Many condensed matter systems are directly mapped to the Coulomb gases. Among them are the XY model~\cite{villain1975j}, the Ashkin-Teller model~\cite{knops1982renormalization}, the $q$-state Potts model~\cite{knops1982renormalization,nienhuis1982analytical}, the antiferromagnetic Pottes model~\cite{den1982critical}, the $O(n)$ model, the frustrated Ising models~\cite{nienhuis1982b}, vortices dynamics in superfluids~\cite{kosterlitz1974jm} and the quantum Hall systems via the plasma analogy of the wave function~\cite{girvin1999quantum}. This correspondence is not restricted to the equilibrium phenomena. Edwards-Wilkinson (EW) model of growth process in the steady state corresponds to GFF which itself corresponds to the Coulomb gas with zero background charge. The EW problem is equivalent to the Laplace equation in the presence of white noise charge~\cite{stanley2012random}. \\
In the present paper we consider the effect of MNPs on the properties of the GFF system in a two-dimensional system. This system is built by solving the Poisson equation in the presence of uncorrelated white-noise Coulomb charges which is related to the EW model in the steady state. The effect of the MNPs is realized by iso-potential clusters which impose some additional boundary conditions. We model the correlations between MNPs by the Ising model with an \textit{artificial temperature} $T$. This temperature is not the real one, but a quantity which controls the mentioned correlations. In fact we map the problem of the formation of MNPs in the isotropic dielectric system to the Ising model for which the temperature plays the role of a control parameter. Such an approach is familiar for other systems, like the position pattern of oxygen atoms in Cu-O planes in YBCO compounds in the normal mode~\cite{najafi2016universality}. The problem is also interesting from the field theory point of view. It is well-known that the GFF corresponds to $c=1$ conformal field theory (CFT), whereas the critical Ising model belongs to $c=\frac{1}{2}$ CFT class. The problem of GFF in the background of Ising-correlated MNP clusters can be viewed as the coupling of $c=1$ and $c=\frac{1}{2}$ CFTs, which is of theoretical interest. It is seen that the behavior of the model at the critical temperature changes significantly with respect to the low-temperature phase, and in the vicinity of the critical (artificial) temperature some power-law behaviors emerge. We find that the system becomes non-Gaussian at (and in the vicinity of) the critical temperature with some exponents different from GFF. Importantly the fractal dimension of the loops (contour lines of the random potential) changes from low-temperature values $D_F^{T=0}=D_F^{\text{GFF}}=\frac{3}{2}$, to the amount $D_F^{T=T_c}=1.4\pm 0.01$ which is closer to the fractal dimension of the fractal dimension of the external perimeter of the 2D critical Ising model $D_F^{\text{Ising}}=\frac{11}{8}$, although the other critical exponents are significantly different. 
The paper has been organized as follows: In the following section, we motivate this study and introduce and describe the model. The numerical methods and details are explored in SEC~\ref{NUMDet}. The results are presented in the section~\ref{results}. We end the paper by a conclusion.

\section{The construction of the problem}
\label{sec:model}
In this section we construct the main idea of the present paper that is the electrostatic potential inside a dielectric media with stochastic iso-potential islands which is further disordered by Coulomb impurities. Noting that in the absence of iso-potential islands, the system corresponds to Gaussian free field (GFF), we can imagine of this problem as the GFF in the background of metallic islands. The spatial pattern of connectedness of MNPs in a dielectric media is generally complex and many internal degrees of freedom play role in the problem, e.g. the cohesive energy, the particle sizes and the effect of the media around. One way out of these complexities retaining the controllable correlations is to map the problem to a model with reduced degrees of freedom. The reduced model should be a two-state model, like the Ising model with a control parameter, i.e. the \textit{artificial} temperature $T$. This mapping reduces the problem degrees of freedom, in favor of simplification, retaining some key parameters which capture the physical processes and allows tunable correlations in the position of MNPs to form the iso-potential islands. Using of the Ising model for the formation pattern of MNP clusters has three benefits: firstly it is a two-state model with tunable correlations, secondly it contains short-range interactions (which is reasonable for the neutral MNPs) and thirdly it is simple for simulation. It is always an important candidate as a reduced model for the systems with two effective states, such as the state of the oxygen atoms in the Cu-O planes of YBCO compounds~\cite{najafi2016universality} and the state of the impermeable rocks in the porous media~\cite{cheraghalizadeh2017mapping}. When the host configuration is made, one can simulate the dynamical model on it, which can be interpreted as the coupling between the dynamical statistical model to the Ising model as the host~\cite{najafi2018coupling}. \\
For the purposes mentioned above, the system is meshed by cells each of which can have one of the two states: empty (not occupied by a MNP) or occupied by a MNP. The system is also supposed to experience the Coulomb disorder which is modeled by white noise, in presence of which the Poisson equation is solved. The spins of the dual Ising model play the role of the field of presence or absence of MNPs. More precisely, if we show the spins by $\sigma$, then $\sigma_i$ equals to $-1$ ($+1$) if a MNP is present (absent) in the $i$th site. The positive-correlation is realized by the ferromagnetic Ising model (positive coupling constant), whereas the negative-correlation is realized by the anti-ferromagnetic one (negative coupling constant). The correlations in the Ising system are controlled by the artificial temperature $T$ which has nothing to do with the temperature of the sample and determines the shape of the iso-potential (metallic) islands. We use this word without 'artificial' throughout this paper, having in mind that we mean the control parameter. The Ising Hamiltonian is:
\begin{equation}
H=-J\sum_{\left\langle i,j\right\rangle}\sigma_i\sigma_j-h\sum_{i}\sigma_i, \ \ \ \ \ \sigma_i=\pm 1
\label{Eq:Ising}
\end{equation}
in which $J$ is the coupling constant ($J>0$ for the ferromagnetic system and $J<0$ for anti-ferromagnetic one), $h$ is the magnetic field which controls the population of the MNPs and $\left\langle i,j \right\rangle$ means that the sites $i$ and $j$ are nearest neighbors. It is worthy to emphasis that we use the Ising model as the metric space and the model is not a magnetic one. For $h=0$ the Ising model is well-known to exhibit a non-zero magnetization per site $M=\left\langle \sigma_i\right\rangle $ at temperatures below a critical temperature $T_c$. In our real system the number of vacant sites (not occupied by the MNPs) is equal to $\frac{1}{2}N(M+1)$ in which $N$ is the total number of sites in the sample in which the MNPs can sit, i.e. $M(T,h)$ is related to the abundance of MNPs. Therefore the population of MNPs and their correlations can be fully tuned by $T$ and $h$. In this paper we set $h$ to zero, i.e. the tendency of the system of being blank or being occupied by MNPs is equal.\\
Although we set $h=0$ throughout this paper, we prefer to mention some points concerning this parameter here. Many local and geometrical properties of the Ising model are known~\cite{delfino2009field,fortunato2002site,fortunato2003critical,najafi2016monte,kose2009label}. There are two transitions in the Ising model: the magnetic (paramagnetic to ferromagnetic) transition and the percolation transition (in which the connected geometrical spin clusters percolate). For the 2D regular Ising model at $h=0$ these two transitions occur simultaneously~\cite{delfino2009field}, although it is not the case for all versions of the Ising model, e.g. for the site-diluted Ising model~\cite{najafi2016monte}. For the details of the percolation transition associated with the critical point of the 2D Ising model see \cite{fortunato2002site} and \cite{fortunato2003critical}. The magnetization has a discontinuity at $h=0$ along the $T<T_c$ line, i.e. for $h=0^+$ and $T<T_c$ we have $M>0$, whereas for the case $h=0^-$ and $T<T_c$ we have $M<0$. The percolation description of the Ising model is as follows: In each $T$ and $h$ the system is composed of some spin clusters. Let us consider only down-spin clusters (corresponding to the presence of MNPs in the main problem), having in mind that the system has the symmetry $h\rightarrow -h$ and $\sigma_i\rightarrow -\sigma_i$. We define $h_{th}(T)$ as the magnetic field threshold below which there is no spanning cluster of the parallel spins and above which some spanning clusters appear. Apparently for $T=0$ all spins align in the same direction and $H_{th}(T=0)=0^+$. Also for $T=\infty$ the spins are uncorrelated and take the up direction with the probability $\frac{1}{2}e^h/\cosh h$. Therefore the percolation threshold $p^{\text{Ising}}_c$ in the case $T\rightarrow\infty$ is:
\begin{equation}
p^{\text{Ising}}_c=\frac{e^{h_{th}(\infty)}}{2\cosh\left( h_{th}(\infty)\right) }.
\end{equation}
By this relation, the infinite temperature Ising model is mapped to the percolation problem. This can be done for an arbitrary temperature with a modified relation.\\
Before describing the problem in this type of media, let us first briefly introduce the standard method of generating GFFs. As mentioned above, the EW model becomes GFF in the steady state. The EW model for a field $V$ is as follows:
\begin{equation}
\partial_t V(\vec{r},t)=\nabla^2V(\vec{r},t)+\eta(\vec{r},t)
\label{Eq:EW}
\end{equation}
in which $\eta(\vec{r},t)$ is a white noise with the properties $\left\langle \eta(\vec{r},t)\right\rangle = 0 $ and $\left\langle \eta(\vec{r},t)\eta(\vec{r}',t')\right\rangle = \zeta \delta^3(\vec{r}-\vec{r}')\delta(t-t') $ and $\zeta$ is the strength of the noise. $V(\vec{r},t)$ is the height field in the EW model, and is the electrostatic potential in our paper (if $t$-independent). In the steady state the problem becomes time-independent ($\partial_tV=0$), acquiring the following form:
\begin{equation}
\nabla^2V(\vec{r})=-\rho(\vec{r})/{\epsilon}
\label{Eq:Poisson}
\end{equation}
which is the Poisson equation of some random Coulomb impurities. In this relation $\rho(\vec{r})$ is the spatial white noise with $\left\langle \rho(\vec{r})\right\rangle = 0 $ and $\left\langle \rho(\vec{r})\rho(\vec{r}')\right\rangle = (n_ia)^2 \delta^3(\vec{r}-\vec{r}')$, $n_i$ is the total density of Coulomb disorder, $a$ is the lattice constant, and $\epsilon$ is the dielectric function of the media around. It is well-known that this model (which corresponds to the steady state of the EW model) in the scaling limit belongs to the $c=1$ conformal filed theory corresponding to the GFF~\cite{francesco1996conformal}. It is also known that the contour lines of this model are described by the Schramm-Loewner evolution (SLE) theory with the diffusivity parameter $\kappa=4$~\cite{Cardy2005Sle}. The fractal dimension of the contour loops $D_f^{\text{GFF}}=\frac{3}{2}$ which is compatible with the relation $D_f=1+\frac{\kappa}{8}$. Before we proceed, it seems necessary to review some features of the scale-invariant 2D random fields and rough surfaces.\\
Let $V(x,y)\equiv V(\mathbf{r})$ be the \textit{height} profile (in this paper the electrostatic potential) of a scale invariant 2D random rough field. The main property of self-affine random fields is their invariance under rescaling \cite{barabasi1995fractal,kirchner2003critical,falconer2004fractal}. The probability distribution function of these fields transform under $\textbf{r}\rightarrow \lambda\textbf{r}$ as follows:
scaling law
\begin{eqnarray}
V(\lambda \mathbf{r}) \stackrel{d}{=} \lambda ^{\alpha} V(\mathbf{r}),
\end{eqnarray}
where the parameter $\alpha$ is \textit{roughness} exponent or the \textit{Hurst} exponent and $\lambda$ is a scaling factor and the symbol $\stackrel{d}{=}$ means the equality of the distributions. Let us denote the Fourier transform of $V(\textbf{r})$ by $V(\textbf{q})$. The distribution of a wide variety of random fields characterized by the toughness exponent $\alpha$ is Gaussian with the form
\begin{eqnarray}
P\left\lbrace  V \right\rbrace \sim \exp \left[ -\frac{k}{2} \int _ 0 ^ {q_0} d \mathbf{q}q^{2(1+\alpha)}V_{\mathbf{q}}V_{-\mathbf{q}} \right],
\end{eqnarray}
where $q_0$ is the momentum cut-off which is of the order of the inverse of the lattice constant~\cite{kondev1995geometrical} and $k$ is some constant. The scale invariance, when combined with the translational, rotational and scale invariance, has many interesting consequences. For example the height-correlation function of $V(\mathbf{r})$, $C(r) \equiv \langle \left[ V(\mathbf{r}+\mathbf{r_0})-V(\mathbf{r_0}) \right]^2 \rangle$ is expected to behave like
\begin{eqnarray}\label{height-corr}
C(r) \sim |\mathbf{r}| ^{2\alpha_l},
\end{eqnarray}
where the parameter $\alpha_l$ is called the local roughness exponent \cite{barabasi1995fractal} and $\left\langle \right\rangle$ denotes the ensemble average. The above equation implies that the second moment of $V(\textbf{q})$ scales with $q$ for small values of $q$, i.e. $S(\mathbf{q})\equiv \langle |V(\mathbf{q})|^2\rangle\sim |\mathbf{q}|^{-2(1+\alpha)}$~\cite{falconer2004fractal} which is obtained from the relation \ref{height-corr}. Another measure to classify the scale invariant profile $V(\mathbf{r})$ is the total variance
\begin{eqnarray}\label{total variance}
W(L)\equiv \langle \left[ V(\mathbf{r}) - \bar{V} \right]^2 \rangle _L \sim L^{2\alpha_g}
\end{eqnarray}
where $\bar{V}=\langle V(\mathbf{r}) \rangle_L$, and $\langle \dots \rangle _L$ means that, the average is taken over $\mathbf{r}$ in a box of size $L$. The parameter $\alpha_g$ is the global roughness exponent. Self-affine surfaces are mono-fractals just if $\alpha_g = \alpha_l = \alpha$ \cite{barabasi1995fractal}. \\
The other test for $V(\textbf{r})$ to be Gaussian is that all of its finite-dimensional probability distribution functions are Gaussian \cite{adler1981geometry}. One of the requirements of this is that its distribution is Gaussian:
\begin{eqnarray}
P( V ) \equiv \frac{1}{\sigma\sqrt{2\pi}}e^{-\frac{V^2}{2\sigma^2}},
\end{eqnarray}
where $\sigma$ is the standard deviation. Another quantity whose moments distributions should be Gaussian is the local curvature which is defined (at position $\mathbf{r}$ and at scale $b$) as~\cite{kondev2000nonlinear}
\begin{eqnarray}\label{local curvature}
C_b(\mathbf{r}) = \sum_{m=1}^M \left[ V(\mathbf{r}+ b\mathbf{e}_m) - V(\mathbf{r}) \right],
\end{eqnarray}
in which the offset directions $\left\lbrace\mathbf{e}_1,\dots,\mathbf{e}_M\right\rbrace$ are a fixed set of vectors whose sum is zero, i.e. $\sum_{m=1}^M \mathbf{e}_m =0$. If the rough surface is Gaussian, then the distribution of the local curvature $P (C_b)$ is Gaussian and the first and all the other odd moments of $C_b$ manifestly vanish since the random field has up/down symmetry $V(\mathbf{r})\longleftrightarrow -V(\mathbf{r})$. Additionally, for Gaussian random fields we have:
\begin{eqnarray}\label{fourth moment}
\frac{\langle C_b^4 \rangle }{\langle C_b^2 \rangle ^2} = 3.
\end{eqnarray}
This relation is an important test for the Gaussian/non-Gaussian character of a random field. \\

All of the analysis presented above are in terms of local variable $V(\textbf{r})$. There is however a non-local view of point in such problems, i.e. the iso-height lines of the profile $V(\mathbf{r})$ at the level set $V(\mathbf{r}) = V_0$ which also show the scaling properties. When we cut the self-affine surface $V(x,y)$ some non-intersecting loops result which come in many shapes and sizes \cite{kondev1995geometrical,kondev2000nonlinear}. We choose $10$ different $V_0$ between maximum and minimum potentials and a \textit{contour loop ensemble} (CLE) is obtained. These geometrical objects are scale invariant and show various power-law behaviors, e.g., their size distribution is characterized by a few power law relations and scaling exponents. The scaling theory of CLEs of self-affine Gaussian fields was introduced in Ref. \cite{kondev1995geometrical} and developed in Ref. \cite{kondev2000nonlinear}. In the following we introduce the various functions and relation, firstly introduced in Ref. \cite{kondev2000nonlinear}. The most important local quantities are $\alpha_l$ and $\alpha_g$. For the non-local quantities the exponent of the distribution functions of loop lengths $l$ ($P(l)$) and the gyration radius of loops $r$ ($P(r)$) are of especial importance. In addition the contour loop ensemble can be characterized through the loop correlation function $G(\mathbf{r})=G(r)$ ($r\equiv |\textbf{r}|$) which is the probability measure of how likely the two points separated by the distance $r$ lie on the same contour. For large $r$s this function scales with $r$ as
\begin{eqnarray}\label{loop correlation function}
G(r) \sim \frac{1}{r^{2x_l}},
\end{eqnarray}
where $x_l$ is the loop correlation exponent. It is believed that the exponent $x_l$ is superuniversal, i.e. for all the known mono-fractal Gaussian random fields in two dimensions this exponent is equal to $\frac{1}{2}$~\cite{kondev1995geometrical,kondev2000nonlinear}. \\
Now consider the probability distribution $P(l,r)$ which is the measure of having contours with length $(l, l + dl)$ and radius $(r, r + dr)$. For the scale invariant CLE, $P(l,r)$ is hypothesized to behave like~\cite{kondev1995geometrical}:
\begin{eqnarray}
P(l,r) \sim l^{-\tau_l -1/D_f} g(l/r^{D_f}),
\end{eqnarray}
where $g$ is a scaling function and the exponents $D_f$ and $\tau_l$ are the fractal dimension and the length distribution exponent, respectively. One also can define the fractal dimension of the loops by the relation $\langle l \rangle \sim r^{\gamma_{lr}}$. By the following straightforward calculation
\begin{eqnarray}\label{loop fractal dimesion}
\langle l \rangle \equiv \frac{\int _0^\infty lP(l,r) dl}{\int _0^\infty P(l,r) dl} \sim r^{D_f},
\end{eqnarray}
we see that $\gamma_{lr}=D_f$. Note also that the probability distribution of contour lengths $P(l)$ is obtained using the relation $P(l) \equiv \int_0^\infty P(l,r)dr \sim l^{-\tau_l}$. It is shown that there are the important scaling relations between the scaling exponents $\alpha$, $D_f$, $\tau_l$ and $x_l$ as follows~\cite{kondev2000nonlinear}:
\begin{eqnarray}\label{hyper1}
D_f(\tau_l -1) = 2-\alpha,
\end{eqnarray}
and
\begin{eqnarray}\label{hyper2}
D_f(\tau_l-3) = 2x_l -2.
\end{eqnarray}\\
Now let us consider the effect of metallic islands. When a MNP configuration is obtained, the Eq.~\ref{Eq:Poisson} is solved for the modified boundary conditions. Apart from the real boundary conditions, the potential inside and on the boundaries of the metallic islands should be constant of the Neumann type. The Ising model (to determine the shape of the MNP islands) is defined on the $L\times L$ square lattice. After obtaining a spin configuration by solving the Eq.~\ref{Eq:Ising} for $h=0$ at a given temperature, the connected spin clusters composed of up spins are identified and the down spins are interpreted as the (occupied) metallic islands. Let us name the on-occupied regions as the \textit{active space}. The random charged impurities are put on the sites in the active space and the Poisson equation is solved with the imposed boundary conditions. In the numerical procedure of solving Eq.~\ref{Eq:Poisson} in the active space the effective coordination number plays an important role which is defined (for the $j$th site) as:
\begin{equation}
z_j\equiv \sum_{i\in \text{neighbors of j}}\delta_{\sigma_i,1}
\end{equation}
in which $\delta$ is the Kronecker delta function. \\
When the iso-potential islands are self-similar ($h=0$ and $T=T_c$ in our problem, in which the Ising model is critical), the problem belongs to the context of the critical phenomena on the fractal systems. This concept was mainly begun by the work of Gefen \textit{et. al.}~\cite{gefen1980critical} in which it was claimed that the critical behavior of the models is tuned by the detail of the topological quantities of the fractal lattice. The cluster fractal dimension, the order of ramification and the connectivity are some examples of these quantities~\cite{gefen1980critical}. This concept can be extended to dilute systems that are fractal in some limits~\cite{cheraghalizadeh2017mapping,najafi2016bak,najafi2016monte,najafi2016water,najafi2018coupling}. There are also some experimental motivations for these models. The examples for the magnetic material in the porous media are \cite{kose2009label,kikura2004thermal,matsuzaki2004real,philip2007enhancement,kim2008magnetic,keng2009colloidal,kikura2007cluster,najafi2016monte}. The critical Ising lattice in this paper plays the role of the fractal lattice on which the GFF is considered. For the case $T\rightarrow 0$ which is a regular lattice, one retrieves the results of the ordinary $c=1$ CFT and also SLE$_{4}$.\\ 
The off-critical temperatures, especially in the close vicinity of the critical temperature are also very important, since they determine the universality class of the model. We will see that in the vicinity of ht critical temperature some power-law behaviors with well-defined critical exponents emerge. \\

In the general theory of critical phenomena, each system in the critical state shows some power-law behaviors for the local and geometrical quantities, i.e. $P(x)\sim x^{-\tau_x}$ ($x=$ the local and geometrical statistical quantities). The estimation of these exponents is a challenging problem, for which a detailed finite-size analysis is required. For finite systems, the finite-size scaling (FSS) theory predicts that~\cite{goldenfeld1992lectures}:
\begin{equation}
P_x(x,L)=L^{-\beta_x}g_x(xL^{-\nu_x}),
\label{eq:FSS}
\end{equation}
in which $g_x$ is a universal function and $\beta_x$ and $\nu_x$ are some exponents that are related by $\tau_x=\frac{\beta_x}{\nu_x}$. The exponent $\nu_x$ determines the cutoff behavior of the probability distribution function. If FSS works, all distributions $ P_x (x,L) $ for various system sizes have to collapse, including their cutoffs. For the critical systems one also expects that $r_{\text{cutoff}}\sim L$, i.e., the cutoff radius should scale linearly with the system size $ L $ ($\nu_r=1$), so that for all observables one gets $ \nu_x = \gamma_{xr} $ \cite{Lubeck1997BTW}. Also we note that there is a hyper-scaling relation between the $\tau$ exponents and the fractal dimensions $\gamma_{x,y}$, which are defined by the relation $x\sim y^{\gamma_{x,y}}$, namely:
\begin{equation}
\gamma_{x,y}=\frac{\tau_y-1}{\tau_x-1}.
\end{equation}
This relation is valid only when the conditional probability function $p(x|y)$ is a function with a very narrow peak for both $x$ and $y$ variables.

\subsection{Numerical methods}\label{NUMDet}

\begin{figure*}
	\centering
	\begin{subfigure}{0.3\textwidth}\includegraphics[width=\textwidth]{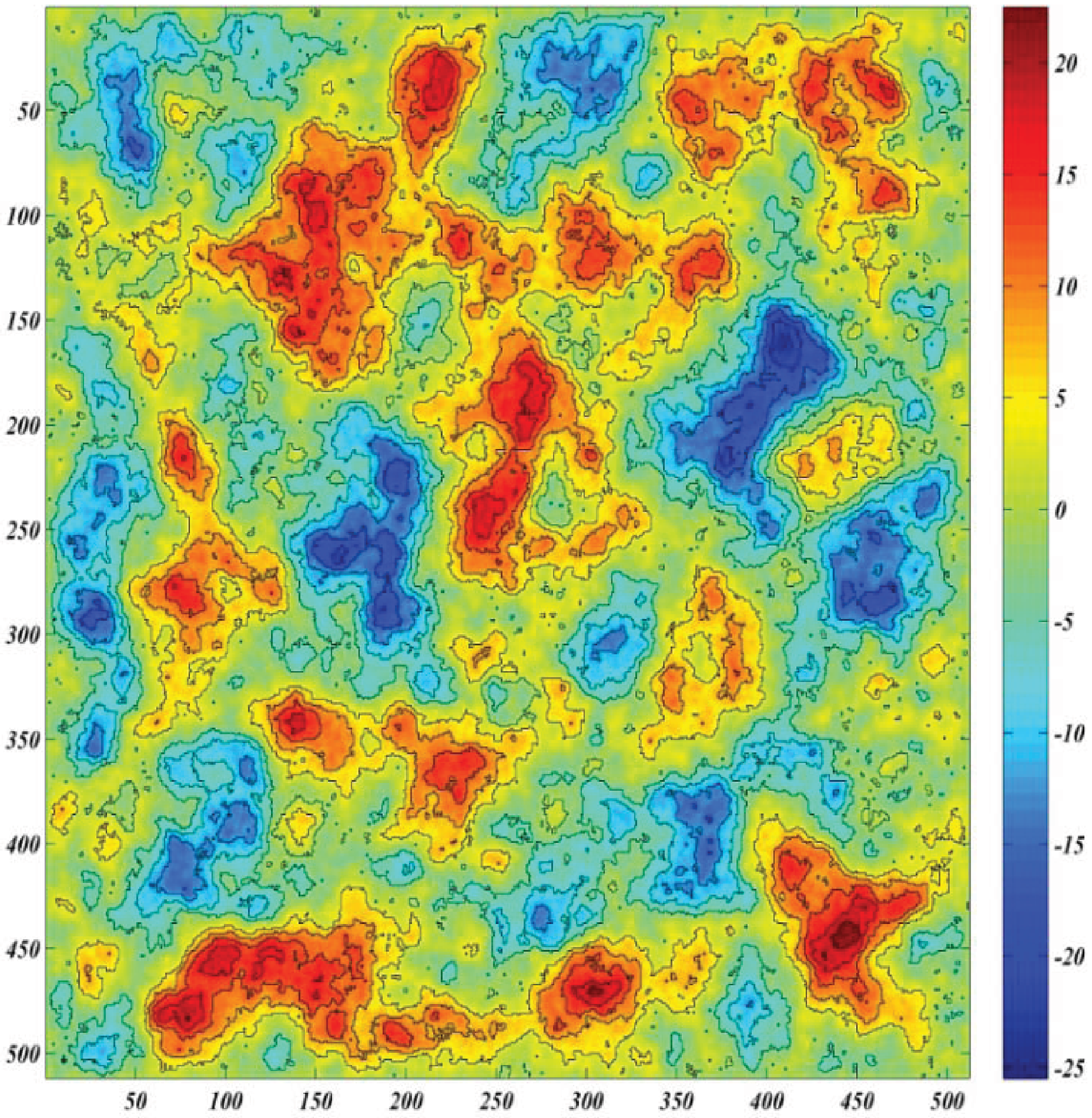}
		\caption{}
		\label{fig:T1}
	\end{subfigure}
	\begin{subfigure}{0.3\textwidth}\includegraphics[width=\textwidth]{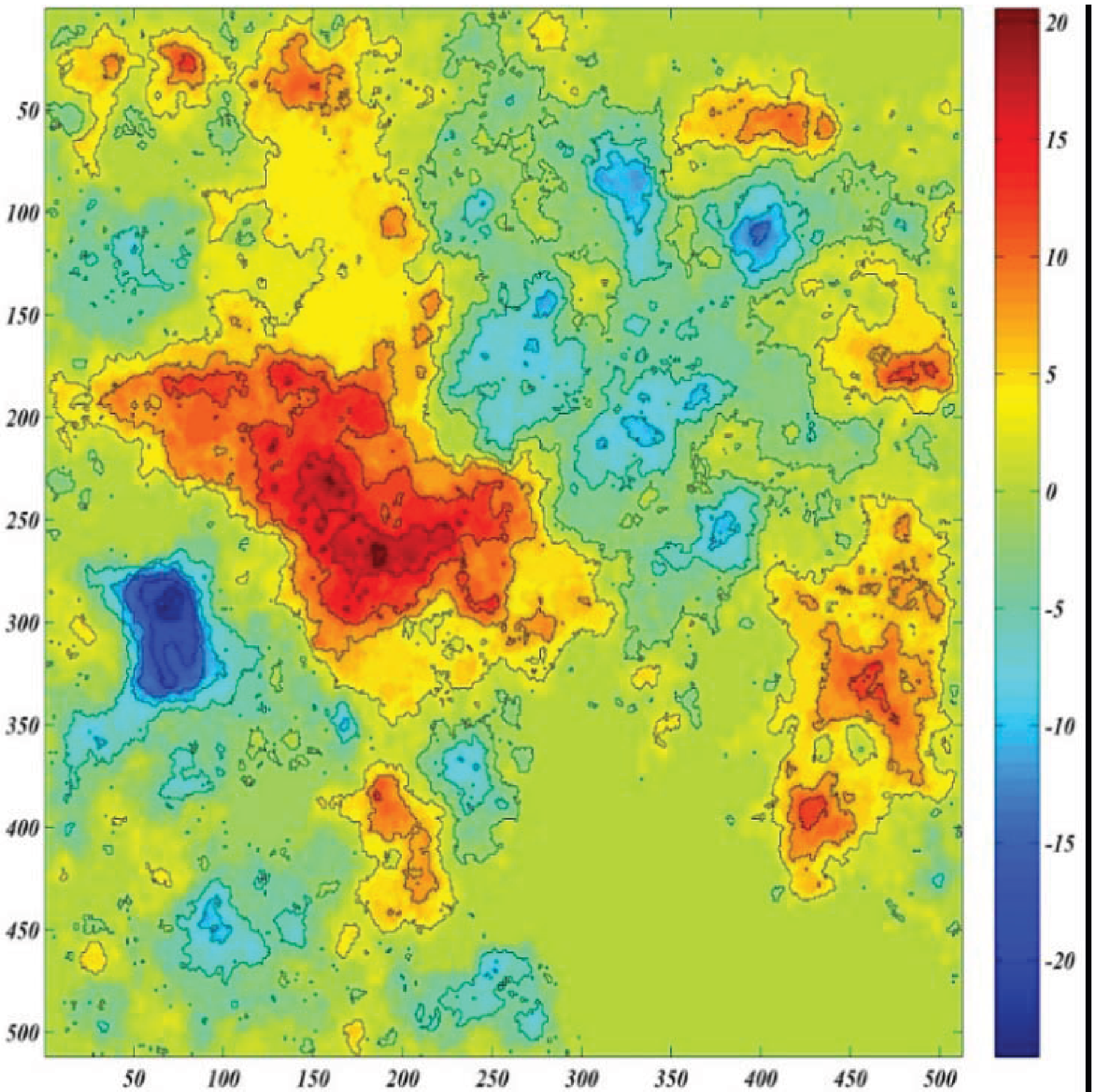}
		\caption{}
		\label{fig:T2}
	\end{subfigure}
	\begin{subfigure}{0.3\textwidth}\includegraphics[width=\textwidth]{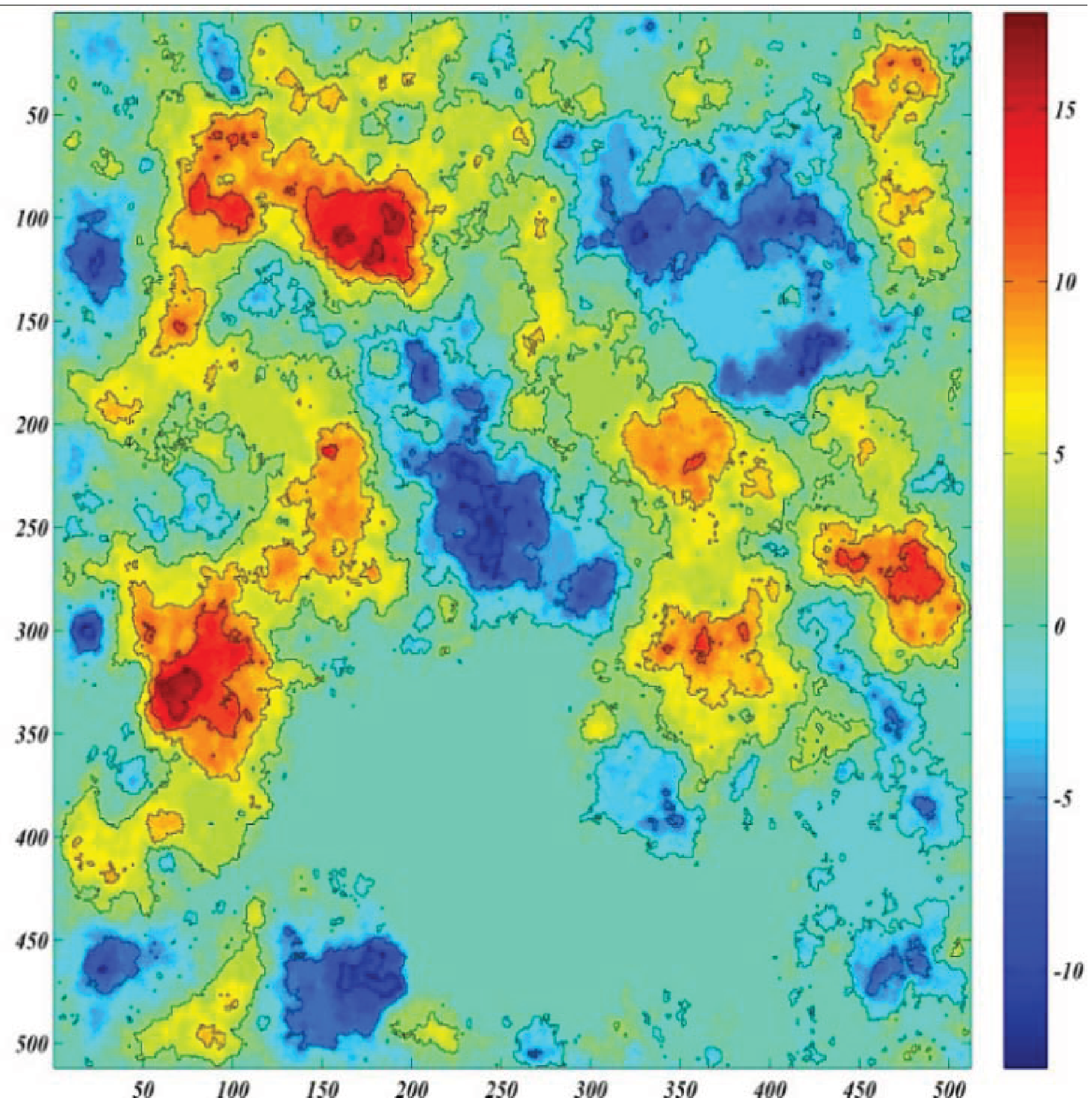}
		\caption{}
		\label{fig:T3}
	\end{subfigure}
	\caption{(Color online) The electrostatic potential for three temperatures: (a) $T=1.868$, (b) $T=2.2681$ and $T=2.2682$.}
	\label{fig:samples}
\end{figure*}

In this paper we have two coupled simulations, one for the Ising model and the other for obtaining the random potential for a particular configuration of Coulomb impurities. The method of simulation in the vicinity of the critical point is important, due to the critical slowing down. To avoid the critical slowing down we have used the Wolff Monte Carlo method to generate Ising samples. Our ensemble averaging contains both random Coulomb impurity averaging as well as the Ising lattice averaging. The Ising independent samples were generated on the lattice sizes $L=64, 128, 256$ and $512$ for each temperature. To make the Ising samples independent, between each successive sampling, we have implied $L^2/3$ random spin flips and let the sample equilibrate by $500L^2$ Monte Carlo steps. The main lattice has been chosen to be square, for which the Ising critical temperature is $T_c\approx 2.269$. Only the samples with temperatures $T\leq T_c$ have been generated, since the spanning clusters (active space) are present only for this case. The temperatures considered in this paper are $T=T_c-\delta t_1\times i$ ($i=1,2,...,5$ and $\delta t_1=0.01$) to obtain the statistics in the close vicinity of the critical temperature (note that the model shows non-trivial power-law behaviors in the vicinity of the critical temperature) and $T=T_c-\delta t_2\times i$ ($i=1,2,...,10$ and $\delta t_2=0.05$) for the more distant temperatures. To obtain the desired samples we have started from the high temperatures ($T>T_c$). For each temperature, $2\times 10^3$ Ising samples were generated and for each Ising sample $10^2$ configurations of Coulomb impurities were tested. For solving the equation~\ref{Eq:Poisson} we have used the finite element method and the self-consistent method with the accuracy parameter $10^{-4}$ was employed. When an electrostatic potential is obtained, we extract the contour lines by $10$ different cuts with the same spacing between maximum and minimum values. It is notable that for each $L=512$ sample (for a given temperature) about $\sim 10^3$ loops were obtained. This means that for each temperature and $L$, something like $10^8$ loops were generated. The Hoshen-Kopelman~\cite{hoshen1976percolation} algorithm has been employed for identifying the clusters in the lattice. Figure \ref{fig:samples} shows samples of the obtained potential for three temperatures: $T=1.868$, $T=2.2681$ and $T=2.2682$. We see that the MNP islands become larger and self-similar as $T$ approaches the critical temperature from bellow. The contour lines for $10$ different cuts have been shown.

\section{Results}\label{results}

For all temperatures, the system show critical behaviors and the random potential pattern $V(\textbf{r})$ is self-similar and scale invariant. However the critical exponents change with the temperature. Two cases can be processed separately: the critical (self-affine) host space ($T=T_c$) and the super-critical one $T<T_c$. As we will see, the critical case is very especial, for which the exponents are substantially different from the low-temperatures. It is interestingly observed that the critical exponents of all observables approach to the ones for the $T=T_c$ lattice in a power-law fashion. The properties of the ordinary GFF is retrieved in the limit $T\rightarrow 0$.

\subsection{local properties}\label{local}

\begin{figure*}
	\centering
	\begin{subfigure}{0.49\textwidth}\includegraphics[width=\textwidth]{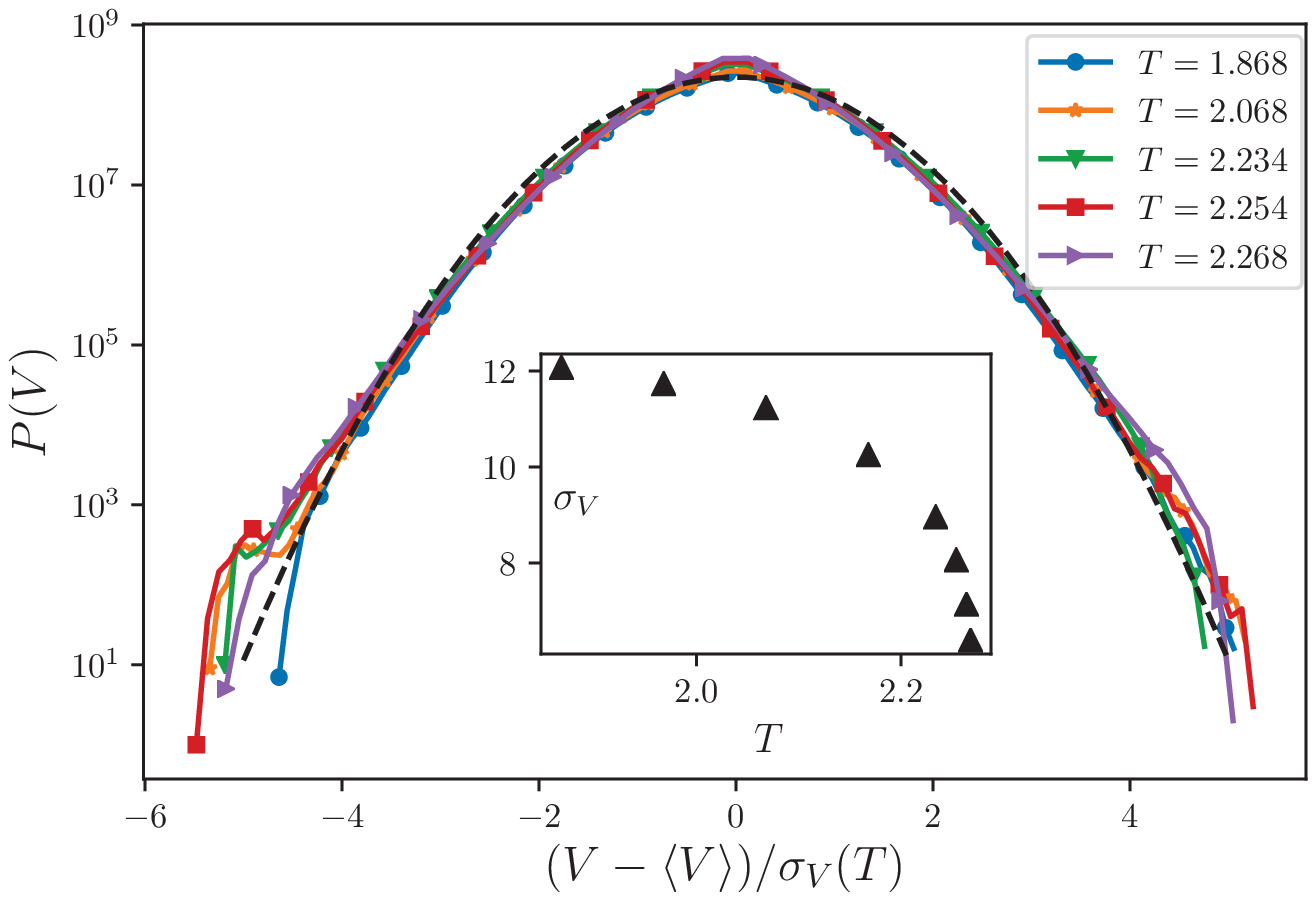}
		\caption{}
		\label{fig:p-h}
	\end{subfigure}
	\begin{subfigure}{0.49\textwidth}\includegraphics[width=\textwidth]{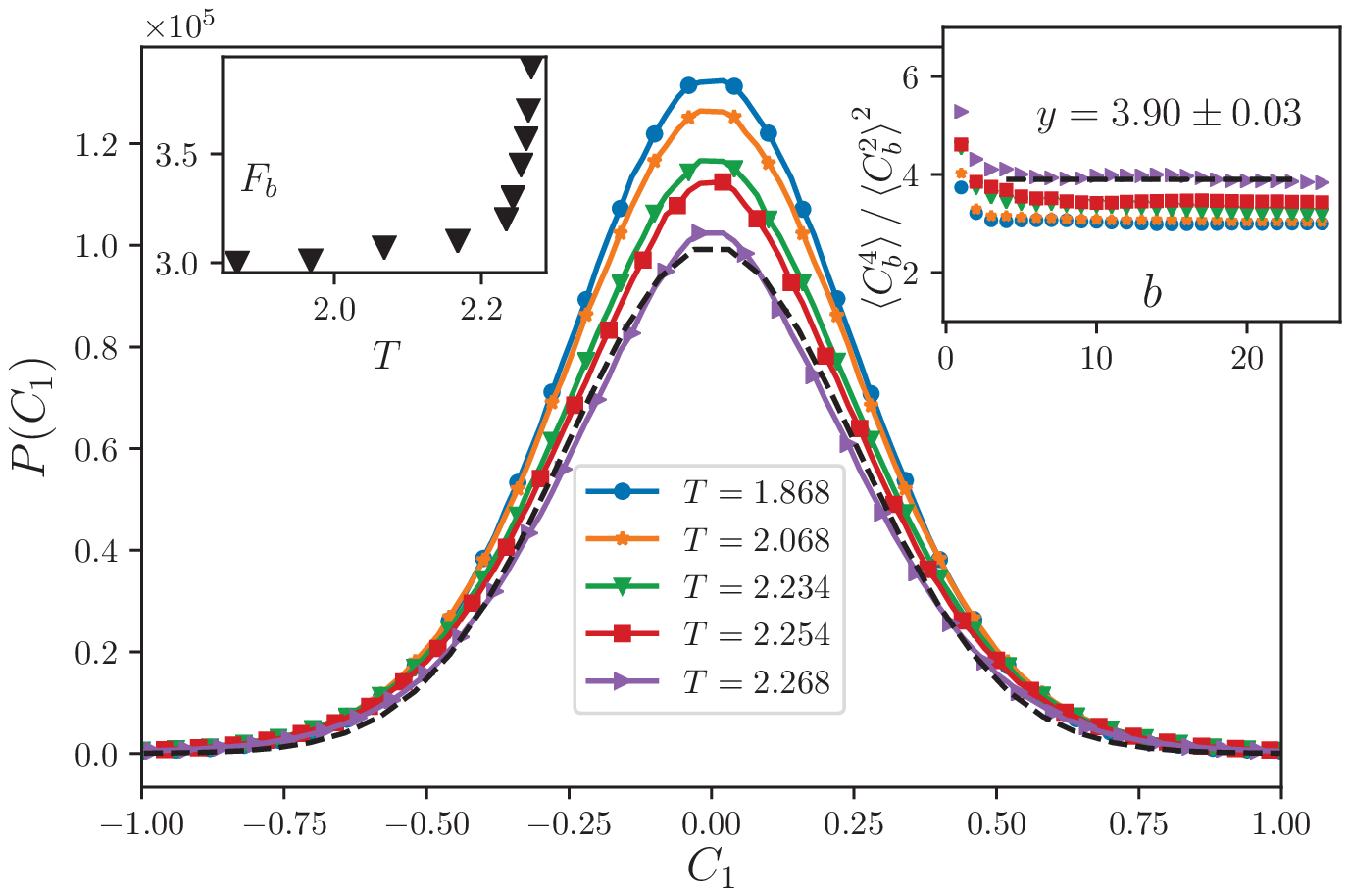}
		\caption{}
		\label{fig:Cb}
	\end{subfigure}
	\begin{subfigure}{0.49\textwidth}\includegraphics[width=\textwidth]{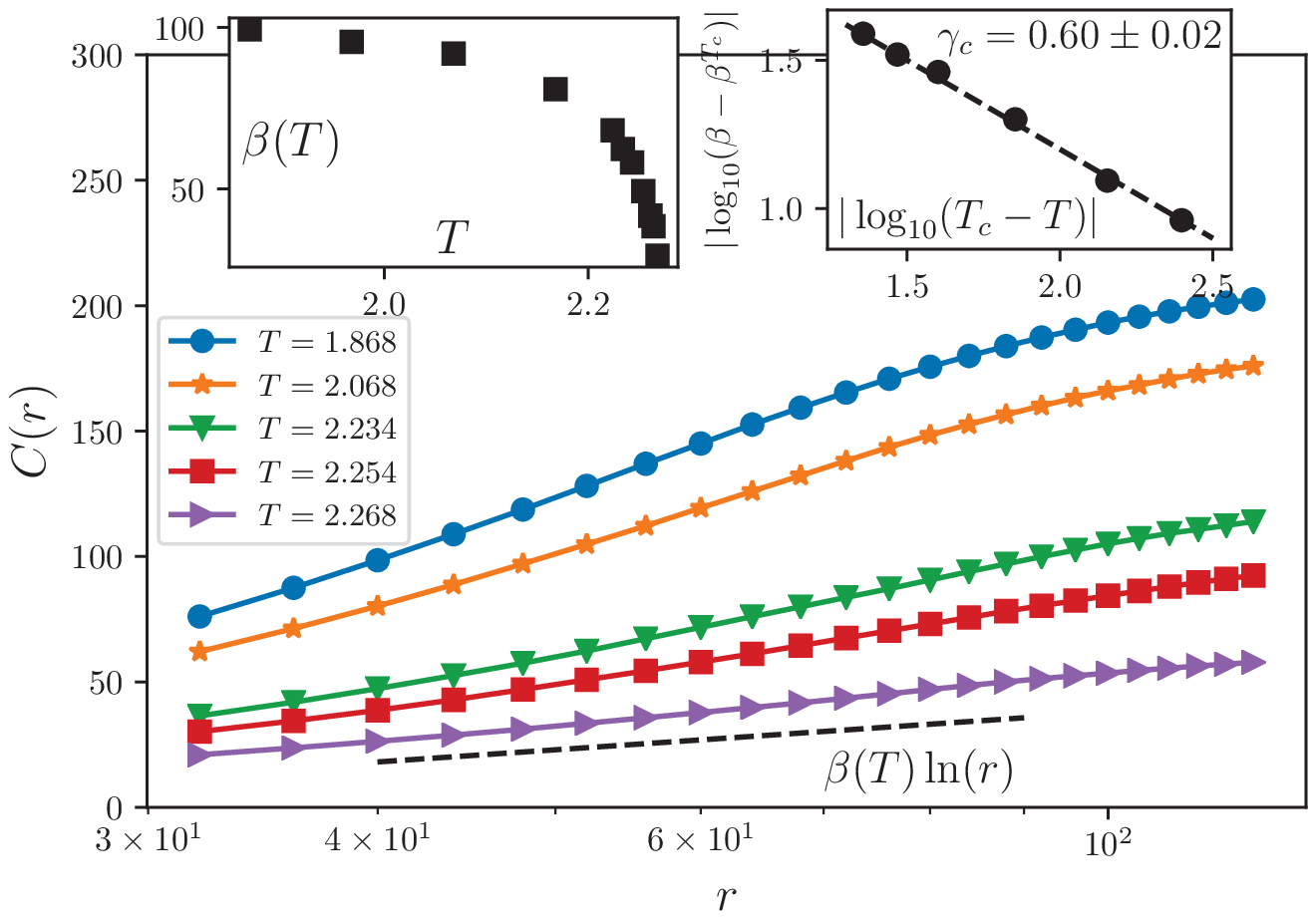}
		\caption{}
		\label{fig:C-r}
	\end{subfigure}
	\begin{subfigure}{0.49\textwidth}\includegraphics[width=\textwidth]{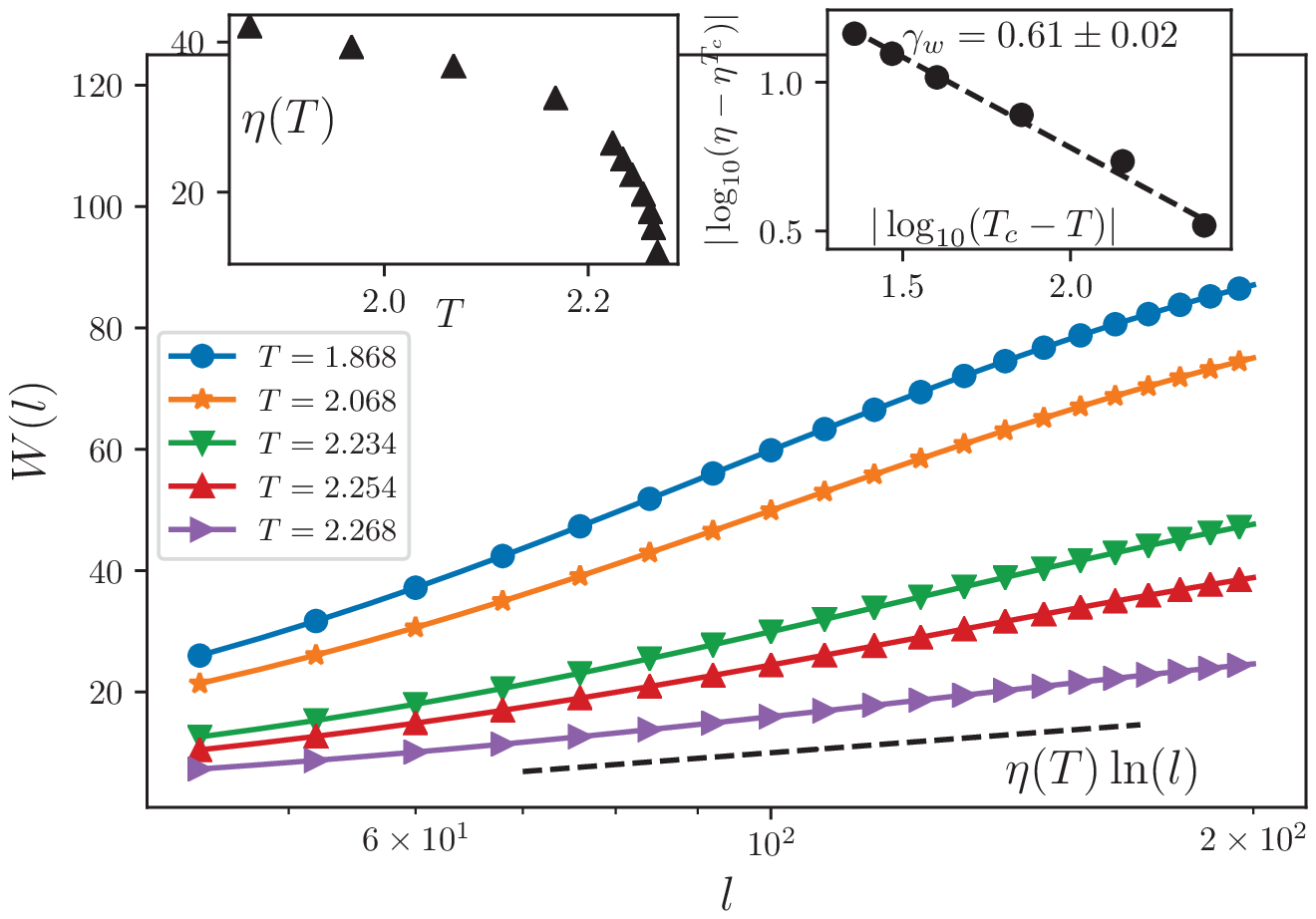}
		\caption{}
		\label{fig:W-l}
	\end{subfigure}
	\caption{(Color online) (a) The distribution of the random potential $P(V)$ with respect to $\left( V-\left\langle V\right\rangle \right) /\sigma_V$ for various temperature for $L=512$. Note that $\left\langle V\right\rangle=0$ for all temperatures. Inset: the variance $\sigma_V$ in terms of temperature. (b) The distribution of the curvature of the random potential $P(C_1)$ for various temperatures. Left inset: $F_b\equiv \frac{\left\langle C_1^4\right\rangle }{\left\langle C_1^2\right\rangle^2}$ in terms of $T$ for $b=20$. Right inset: $F_b$ in terms of $b$. Logarithmic behavior of (c) $C(r)$ and (d) $W(l)$ with the power-law behavior of the proportionality constants $\beta(T)$ and $\eta(T)$ respectively.}
	\label{fig:Off-Tc}
\end{figure*}

One of the important issues in the random field surfaces is the Gaussian/non-Gaussian properties. The question whether the distribution of a random field is Gaussian can addressed directly by calculating the distribution function of the field itself (here $P(V)$), and the corresponding curvature field $P(C_b)$ as explained in the previous section. The fact that these functions are Gaussian is a necessary, but not sufficient condition for Gaussian random fields. These functions have been shown in Figs.~\ref{fig:p-h} and~\ref{fig:Cb}, that are properly fitted to Gaussian distribution. Although the mean potential $\left\langle V\right\rangle $ is zero for all temperatures, the variance of $P(V)$ ($\sigma_V$) decreases with increasing $T$, showing that the distribution of the potential is sharpest in $T=T_c$ and the statistical fluctuations is larger for lower temperatures and correspondingly smaller numbers of MNPs. Therefore growing MNPs in a dielectric media reduces the statistical fluctuations of the electrostatic potential in the presence of Coulomb impurities. This may have applications in the Schrodinger equation in these systems with disorder potentials. \\
For a Gaussian distribution, $F_b\equiv\left\langle C_b^4\right\rangle /\left\langle C_b^2\right\rangle^2=3$, which is the case for low temperatures. As is shown in the insets of Fig.~\ref{fig:Cb}, for the higher temperatures however, this function varies and reaches $\sim 3.9$ as $T\rightarrow T_c$. This shows that the system deviates from the Gaussian form.\\

The most important local quantity in rough surfaces is the $\alpha$ exponent. Although the system become non-Gaussian for temperatures near the critical temperature, the $\alpha$ exponents are well-defined. In the GFF ($T=0$ in this paper) it is well-known that $\alpha=0$ and $C(r)$ changes logarithmically with respect to $r$ as is evident in Fig.~\ref{fig:C-r}. The MNP islands do not change this behavior and for all considered temperatures this behavior has been observed with different proportionality constant, i.e. $C(r)=\beta_T\log(r)$ in which $\beta_T$ is the proportionality constant which is $T$-dependent. As is seen in the inset of Fig.~\ref{fig:C-r}, this dependence is power-law ($\left| \beta(T)-\beta(T_c)\right| \sim t^{\gamma_c}$, $t\equiv \frac{\left| T-T_c\right| }{T_c}$) with the exponent $\gamma_c=0.60\pm 0.02$. The same is seen for the roughness $W(L)$ which changes logarithmically with $L$, i.e. $W(L)=\eta(T)\log(L)$ in which $\left| \eta(T)-\eta(T_c)\right\| \sim t^{\gamma_w}$ with $\gamma_w=0.61 \pm 0.02$. We see that the effect of MNPs is just changing the proportionality constant, and $\alpha_l=\alpha_g=0$ for all temperatures. The results have been gathered in TABLE~\ref{tab:local-exponents}.\\

\begin{table}
	\begin{tabular}{c|c|c}
		\hline Exponent & Definition & value \\
		\hline $\alpha_l$ & $C(r)\sim r^{\alpha_l}$ & $0$ \\
		\hline $\alpha_g$ & $W(L)\sim L^{\alpha_g}$ & $0$ \\
		\hline $\gamma_c$ & $\left| \beta_T-\beta_{T_c}\right| \sim t^{\gamma_c}$ & $0.60\pm 0.02$ \\
		\hline $\gamma_w$ & $\left| \eta_T-\eta_{T_c}\right| \sim t^{\gamma_w}$ & $0.61\pm 0.02$ \\
		\hline
	\end{tabular}
	\caption{The critical exponents of the local quantities.}
	\label{tab:local-exponents}
\end{table}

We conclude that although the MNPs sharpen the distribution function of the random potential, they do not change the logarithmic behavior of the two-point correlation function $C(r)$ as well as the local roughness $W(L)$. The analysis on the global properties of the model is necessary to determine the change of the universality class of the model with respect to $T=0$ GFF.

\subsection{geometrical properties}

\begin{figure*}
	\centering
	\begin{subfigure}{0.49\textwidth}\includegraphics[width=\textwidth]{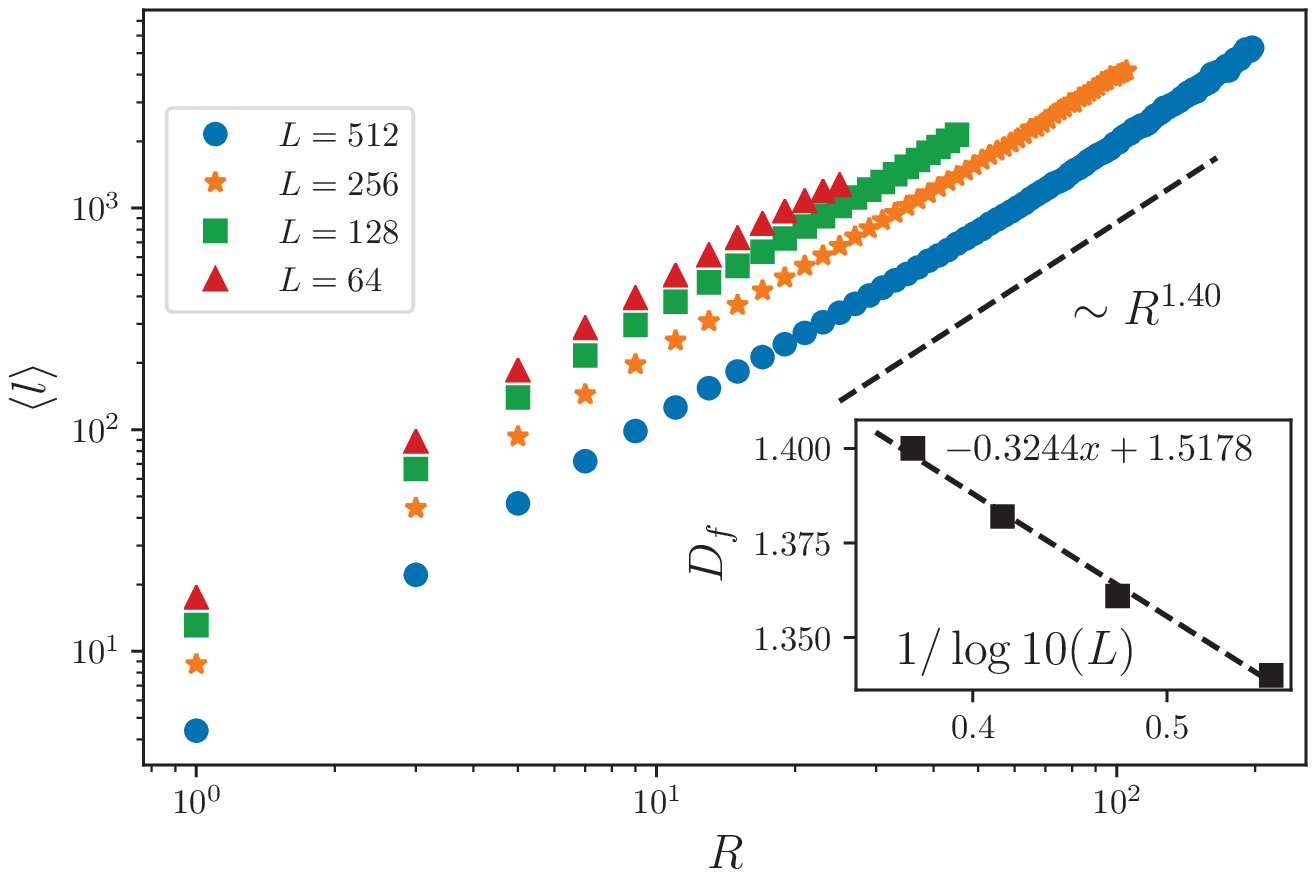}
		\caption{}
		\label{fig:L-Df}
	\end{subfigure}
	\begin{subfigure}{0.49\textwidth}\includegraphics[width=\textwidth]{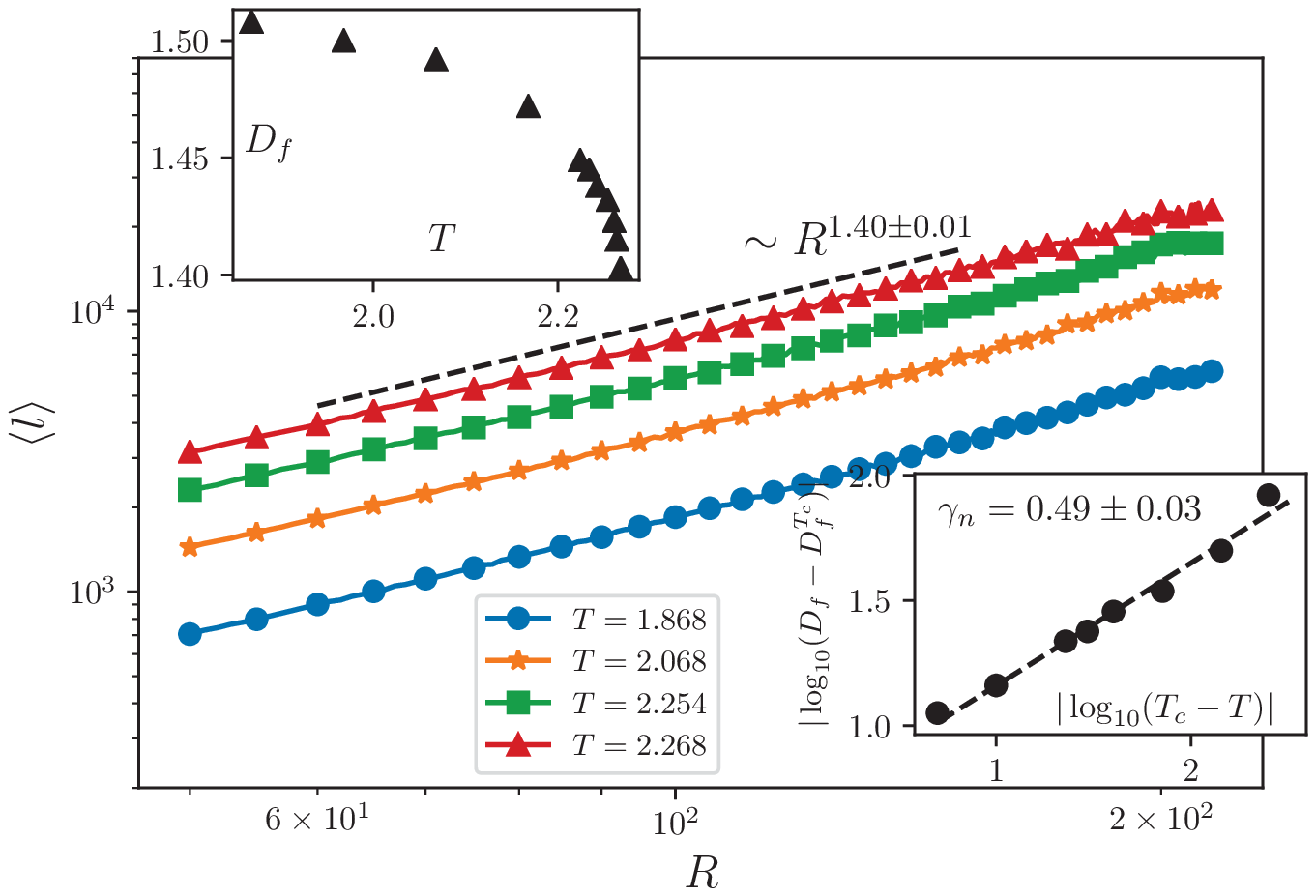}
		\caption{}
		\label{fig:Df}
	\end{subfigure}
	
	\caption{(Color online) (a) The fractal dimension defined by $\left\langle l\right\rangle\sim R^{D_f} $ (a) in terms of system size ($L$) for $T=T_c$ and (b) in terms of $T$ for $L=512$. The inset of (b): the power-law variation of the fractal dimension with respect to the temperature.}
	\label{fig:geometrical1}
\end{figure*}

The local features of the critical models imply some non-local properties, which make the problem worthy to be investigated from the geometrical point of view. This helps to distinguish more precisely the universality class of the model in hand. The critical exponents of the distribution functions are examples of these exponents. There are also some key geometrical exponents which are uniquely dedicated to universality classes and can be interpreted as the representative of that class. The fractal dimension of the contour loops of the critical model is one of these quantities. According to the SLE theory, the 2D critical models are classified using the diffusivity parameter $\kappa$. The interfaces of such models are described by SLE$_{\kappa}$. The fractal dimension $D_f$ of the interfaces of the model is related to the diffusivity parameter by $D_f=1+\frac{\kappa}{8}$, which shows the importance of this exponent.

\begin{figure*}
	\centering
	\begin{subfigure}{0.49\textwidth}\includegraphics[width=\textwidth]{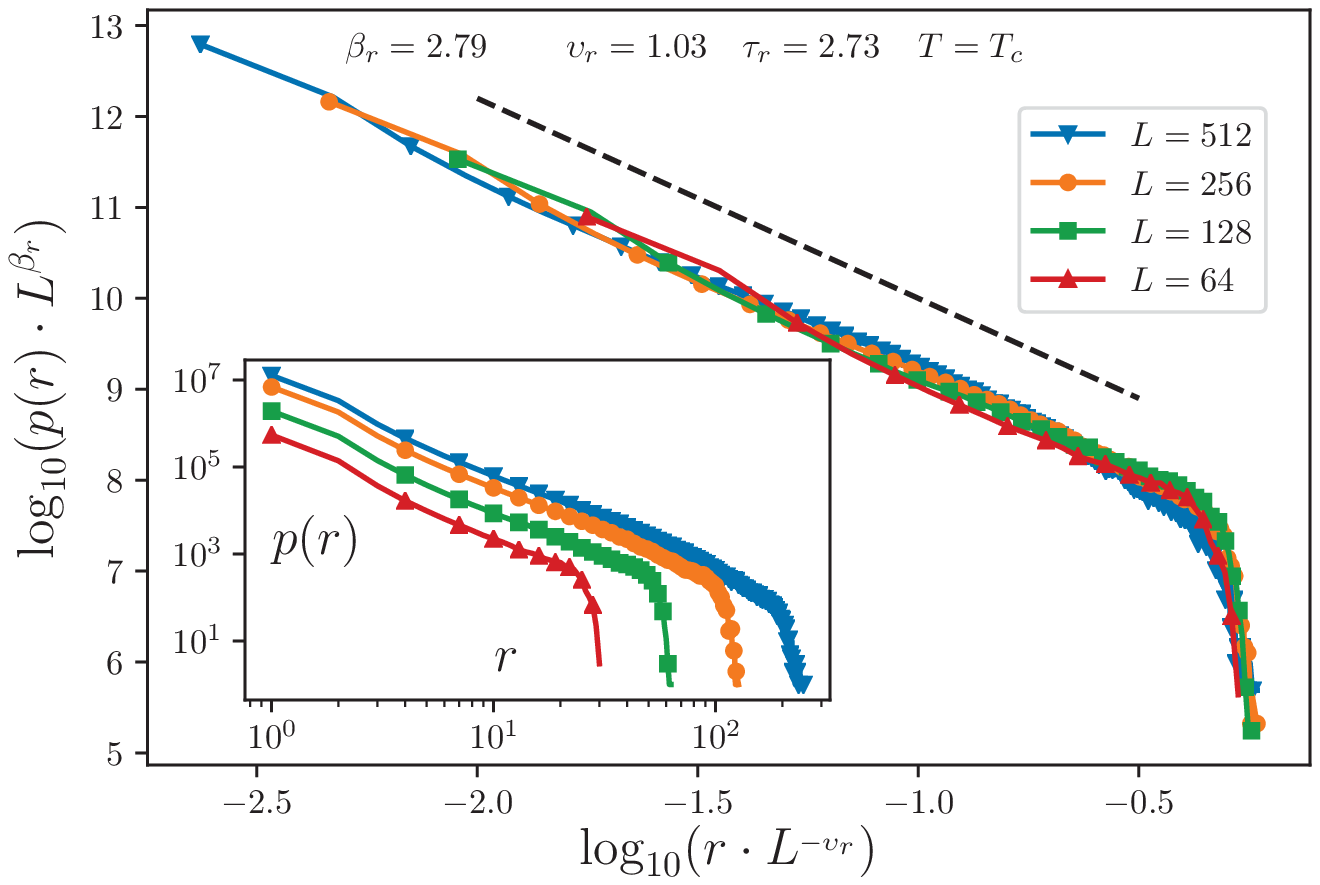}
		\caption{}
		\label{fig:p-r-L}
	\end{subfigure}
	\begin{subfigure}{0.49\textwidth}\includegraphics[width=\textwidth]{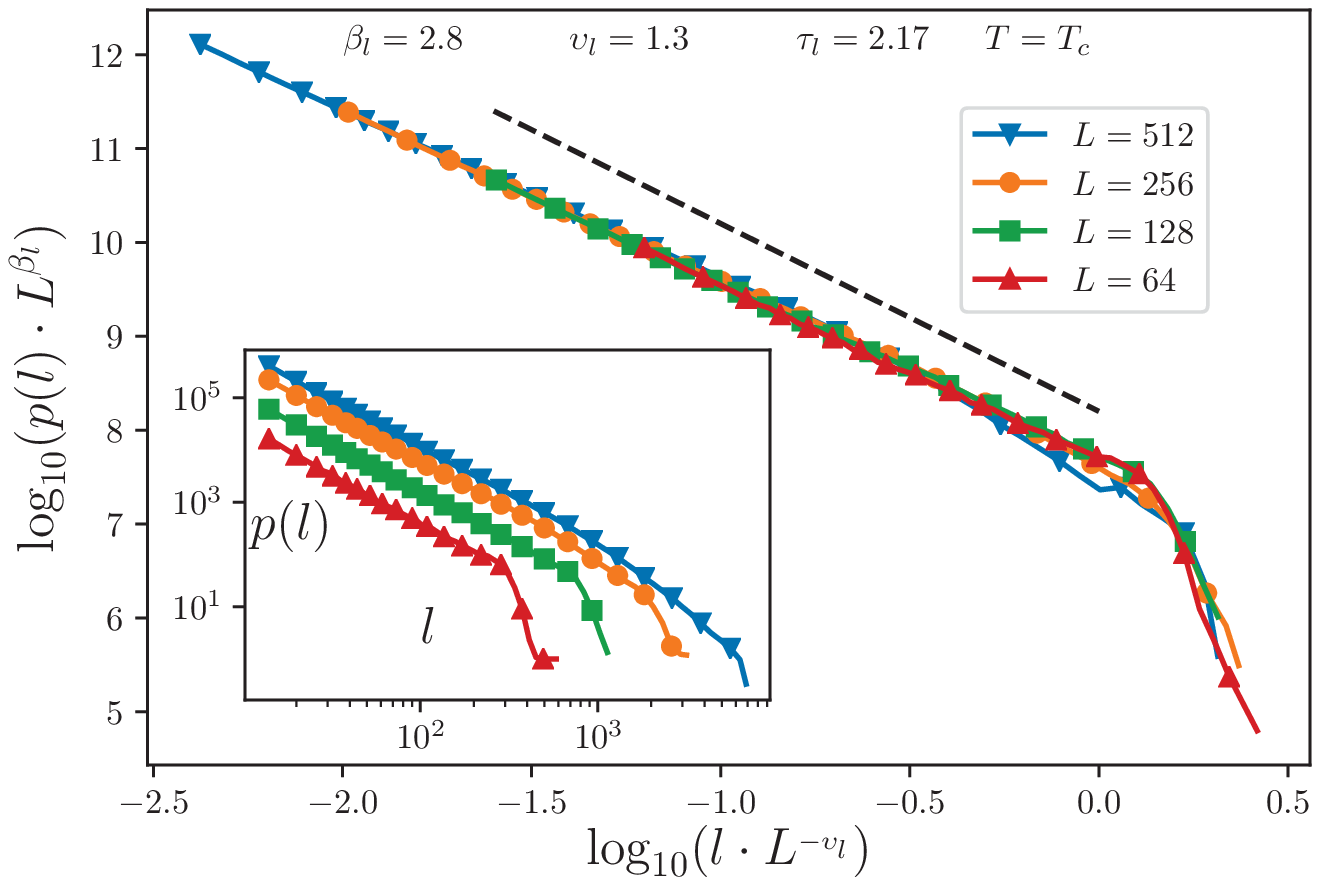}
		\caption{}
		\label{fig:p-l-L}
	\end{subfigure}
	\begin{subfigure}{0.49\textwidth}\includegraphics[width=\textwidth]{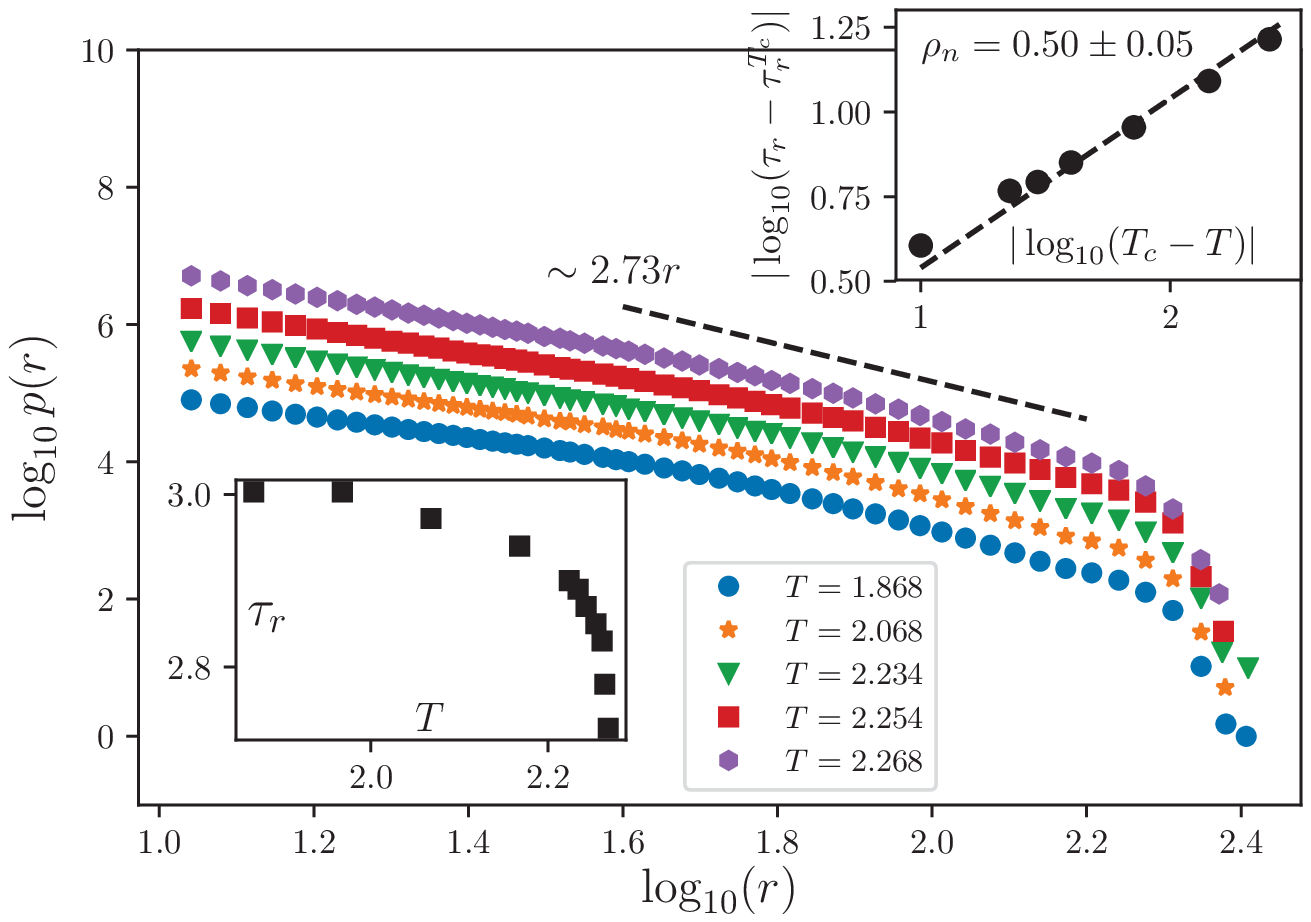}
		\caption{}
		\label{fig:Tr}
	\end{subfigure}
	\begin{subfigure}{0.49\textwidth}\includegraphics[width=\textwidth]{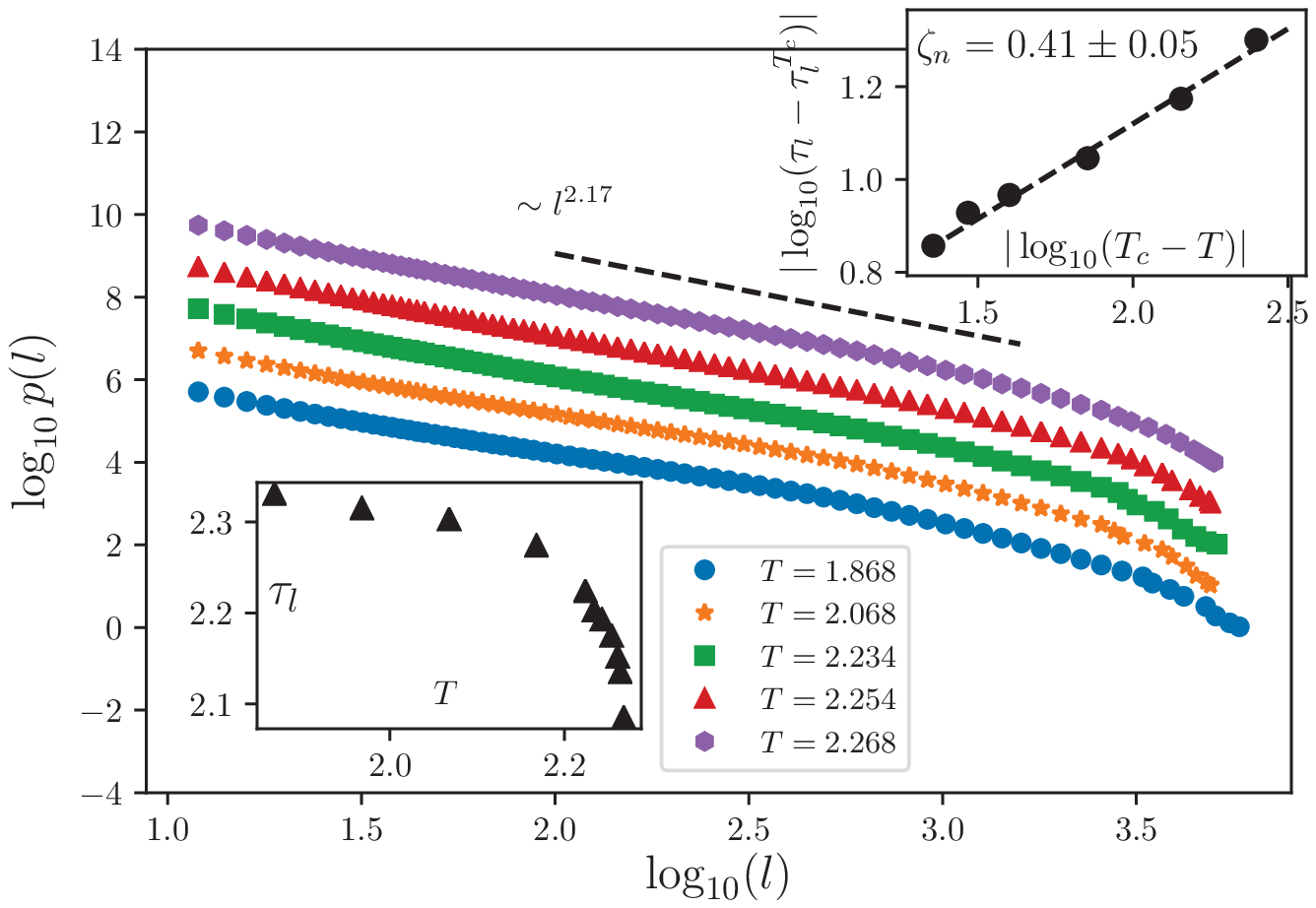}
		\caption{}
		\label{fig:Tl}
	\end{subfigure}
	\begin{subfigure}{0.49\textwidth}\includegraphics[width=\textwidth]{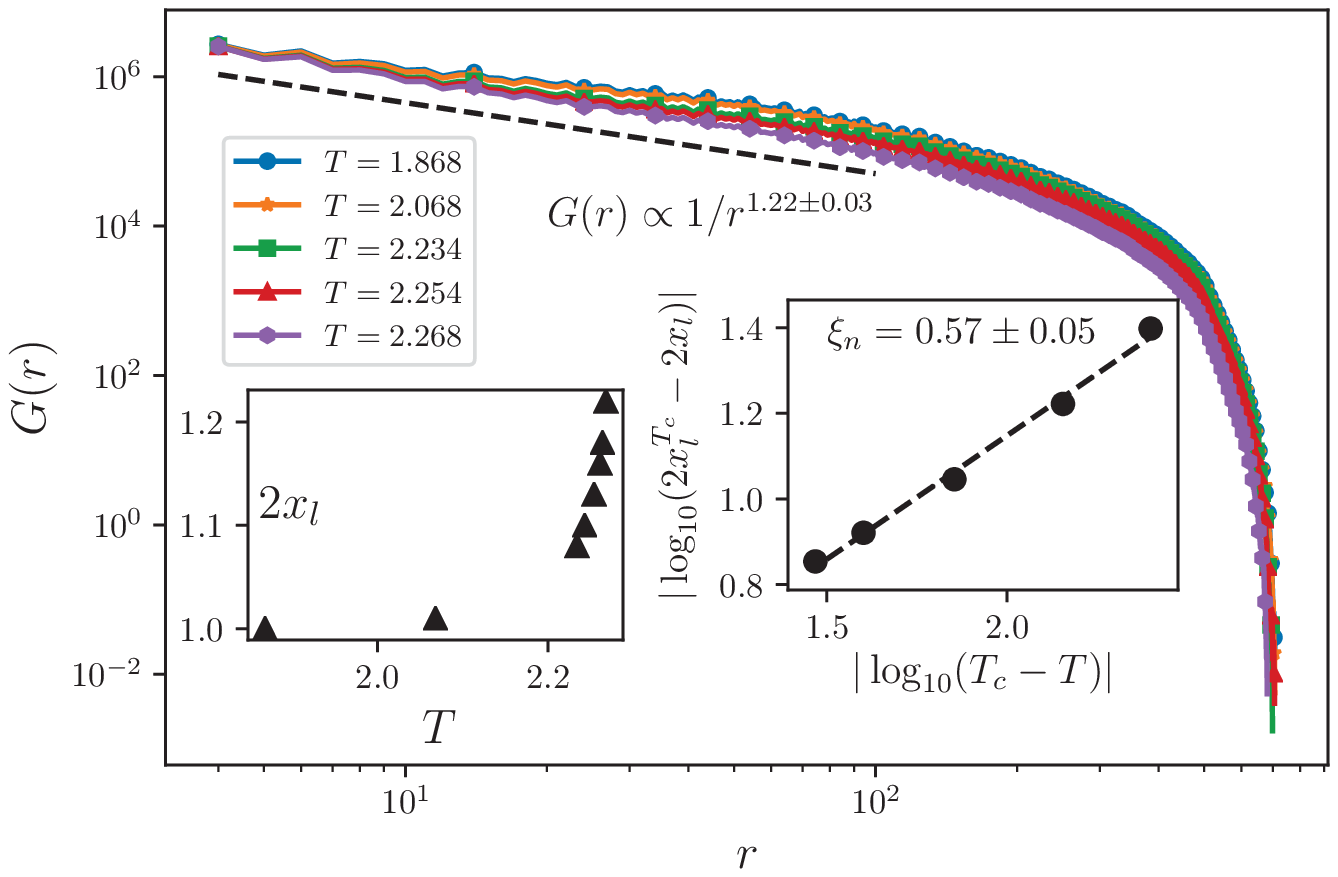}
		\caption{}
		\label{fig:G-r}
	\end{subfigure}
	\caption{(Color online) The data collapse analysis of the distribution function of (a) the gyration radius $r$, (b) the loop length $l$. The temperature dependence of (c) $p(r)$, (d) $p(l)$. Insets: the power-law behavior of $\tau_r$ and $\tau_l$ respectively. (e) The loop Green function $G(r)$ with the exponent $x_l$ which varies with $T$ in a power-law fashion.}
	\label{fig:geometrical2}
\end{figure*}

The fractal dimension of the contour loops at the critical temperature $T=T_c$ for various lattice sizes has been shown in Fig.~\ref{fig:geometrical1}. As seen in this figure $D_f(T_c)$ changes with $1/\log(L)$. For $L=512$ the fractal dimension is obtained $1.40\pm 0.02$. This exponent differs from (is lower than) the fractal dimension of contour lines of GFF $6.6\%$ ($D_f^{\text{GFF}}=D_f^{T=0}=\frac{3}{2}$~\cite{kondev2000nonlinear}) and $7.3\%$ from the fractal dimension of the external perimeter of the spin clusters of the critical 2D Ising model ($D_f^{\text{Ising}}=\frac{11}{8}$~\cite{cardy2005sle}). The power-law behavior of $D_f(T)-D_f(T_c)$ in terms of $t$ is shown in Fig.~\ref{fig:Df} with the exponent $\gamma_n$. We have obtained the exponent $\gamma_n$ equals $0.49\pm 0.03$ whose closest fractional value is $\frac{1}{2}$. This relation is held for all lattice sizes considered in this paper. Noting that the correlation length of the 2D Ising model $\xi$ scales with temperature as $\xi\sim |T-T_c|^{-1}$, one finds the following scaling relation:
\begin{equation}
D_F(T)-D_F(T_c)\sim \frac{1}{\sqrt{\xi}}.
\end{equation}
This $\frac{1}{2}$ exponent is expected to be the most important off-critical geometrical exponent in the problem which characterizes the universality class of the model in hand. This exponent is interestingly the same as the exponent for the self-avoiding walk on the Ising-correlated percolation lattice~\cite{cheraghalizadeh2018self}, although the fractal dimension at $T=T_c$ is different.\\

The exponents $\tau_r$ and $\tau_l$ have been calculated in Figs.~\ref{fig:p-r-L} and \ref{fig:p-l-L} for $T=T_c$ for various lattice sizes, and in Figs.~\ref{fig:Tr} and \ref{fig:Tl} for all temperatures for $L=512$. The data collapse analysis shows that the finite size hypothesis works and $\tau_r(T_c)=2.73\pm 0.1$ and $\tau_l(T_c)=2.17$. These values should be compared with the GFF exponents, i.e. $\tau_r^{\text{GFF}}=\tau_r^{T=0}=3$ and $\tau_l^{\text{GFF}}=\tau_l^{T=0}=\frac{7}{3}$. These exponents (at $T=T_c$) have been gathered in TABLE~\ref{tab:global-exponents1}. The obtained exponents are meaningfully different from the corresponding values for GFF and Ising model, as shown in this table.\\

\begin{table}
	\begin{tabular}{c|c|c|c|c}
		\hline  & $\beta$ & $\nu$ & $\beta/\nu$ & $\tau$ (slope) \\
		\hline $r$ & $2.8\pm 0.1$ & $1.03\pm 0.02$ & $2.7\pm 0.1$ & $2.7\pm 0.1$ \\
	    \hline $r$:(GFF,Ising) & -- & -- & -- & ($3$\cite{kondev2000nonlinear},$3.4\pm 0.1$\cite{najafi2015observation})\\
		\hline $l$ & $2.8\pm 0.1$ & $1.3\pm 0.05$ & $2.15\pm 0.05$ & $2.2\pm 0.1$\\
		\hline $l$:(GFF,Ising) & -- & -- & -- & ($\frac{7}{3}$\cite{kondev2000nonlinear},$\frac{27}{11}$\cite{najafi2015observation})\\
		\hline
	\end{tabular}
	\caption{The critical exponents of the global quantities ($\tau_r$ and $\tau_l$) at $T=T_c$, resulting from the data collapse and the slope of the graph for $L=512$. The corresponding exponents for the GFF and the critical Ising models have been shown for comparison.}
	\label{tab:global-exponents1}
\end{table}

The result that the MNPs lower both $\tau_r$ and $\tau_l$, along with reducing the fractal dimension of curves, show that the MNPs soften the random potential and correspondingly the iso-potential contours. This is compatible with the result of the previous subsection, i.e. the MNPs reduce the fluctuations of the random potential. The fact that more fluctuating potential results to more rough and twisted iso-potential lines, leads one to conclude that the MNPs soften the fluctuations of the random potential. This softening occurs in a power-law form which have been shown in Fig.~\ref{fig:Tr} (for $\tau_r$) and Fig.~\ref{fig:Tl} (for $\tau_l$). It is very interesting that the resulting exponents are more or less equal to $\gamma_n\simeq 0.5$. The results have been shown in TABLE~\ref{tab:global-exponents2}.\\

\begin{table}
	\begin{tabular}{c|c|c}
		\hline Exponent & Definition & value \\
		\hline $\gamma_n$ & $\left| D_f(T)-D_f(T_c)\right| \sim t^{\gamma_n}$ & $0.49\pm 0.03$ \\
		\hline $\rho_n$ & $\left| \tau_r(T)-\tau_r(T_c)\right| \sim t^{\rho_n}$ & $0.5\pm 0.05$ \\
		\hline $\zeta_n$ & $\left| \tau_l(T)-\tau_l(T_c)\right| \sim t^{\zeta_n}$ & $0.41\pm 0.05$ \\
		\hline $\xi_n$ & $\left| x_l(T)-x_l(T_c)\right| \sim t^{\xi_n}$ & $0.57\pm 0.05$ \\
		\hline
	\end{tabular}
	\caption{The critical exponents of the global quantities away from the critical temperature.}
	\label{tab:global-exponents2}
\end{table}

It is well-known that $x_l$ is superuniversal for all known Gaussian random field and is equal to $\frac{1}{2}$ (as stated in the previous section). This exponent changes considerably for the temperatures in the vicinity of $T_c$. This confirms that the random potentials for high temperatures (close to the critical temperature) are non-Gaussian in accordance with the results of the previous sub-section. This has been shown in Fig.~\ref{fig:G-r} in which $G(r)$ has been plotted for various temperatures for $L=512$. It is seen that $x_l(T)-x_l(T_c)$ varies in a power-law form with respect to $t$, with the exponent $\xi_n$ whose numerical value is approximated to be $0.57\pm 0.05$. 

\section*{Discussion and Conclusion}
\label{sec:conc}

In this paper we have considered the effects of metallic nano-particles (MNPs) on the electrostatic potential of a 2D dielectric media which is disordered by white-noise Coulomb impurities. The MNPs are assumed to be correlated, whose position configuration is modeled by the Ising model with an artificial temperature that controls the correlations. The problem can be mapped to the Edwards-Wilkinson (EW) model of surface growth in the steady state, or the Gaussian free field (GFF) in the background of iso-height islands (which are the iso-potential MNP islands in the main problem). Therefore this problem can be viewed as a scale-invariant random surface with the critical exponents which are $T$-dependent (correlation-dependent). This brings the effect of iso-height islands and the correlation between them into the Gaussian random fields, which is GFF in this paper. \\
Two kind of observables have been processed: local and global quantities. The height correlation function $C(r)$ and the total variance (roughness function) $W(L)$ remain logarithmic for all considered temperatures ($T$), just like the GFF for which $\alpha_l=\alpha_g=0$. The proportionality constants of these functions however change with $T$ in a power-law fashion, which has its roots in the criticality of the system. The exponents of these power-law behaviors are nearly the same which have been reported in the text. This power-law behavior is not limited to these local quantities, but also geometrical quantities show the same behaviors.\\
The investigations on the fractal dimension of iso-height lines $D_f$, as well as the critical exponents of the distribution function of the gyration radius $\tau_r$ and loop lengths $\tau_l$ of contour loop ensemble (CLE) show that MNPs soften the random potential and correspondingly the iso-potential contours. This is compatible with the result of the local quantities, i.e. the MNPs reduce the fluctuations of the random potential as has been observed in its distribution. In the off-critical regime ($T<T_c$), close to the critical temperatures, some power-law behaviors have been observed for $D_f$, $\tau_r$, $\tau_l$ and $x_l$ (the critical exponent of the loop Green function $G(r)$). The critical exponents have been gathered in TABLE~\ref{tab:global-exponents2}. The important geometrical exponent is the $\frac{1}{2}$ exponent in $D_f(T)-D_f(T_c)\sim \frac{1}{\sqrt{\xi(T)}}$ in which $\xi(T)$ is the spin correlation length which is $T$-dependent. \\
The behavior of the fractal dimension in the vicinity of a critical point is always important, and reveals some features of the universality class. In many critical systems, the off-critical dual model in the scaling limit are massive models. The examples are the dissipative sandpile model which is equivalent to massive ghost model, and the off-critical Ising model which corresponds to the massive Majorana Fermions. It has been observed that in some cases, the fractal dimension of the loops (as the interfaces of the model) is proportional to the square root of the \textit{mass} parameter. The example is that $D_f^m-D_f^{m=0}\sim m^{\frac{1}{2}}$ for the dissipative avalanches, in which $D_f^m$ is the fractal dimension of the external frontier of the avalanches~\cite{najafi2012avalanche}. The similar behavior is seen for the fractal dimension of the self-avoiding walks on the off-critical Ising-correlated lattice, i.e. $D_f(T)-D_f(T_c)\sim m^{\frac{1}{2}}$ in which $m\equiv t$ is the effective mass of the dual Majorana Fermions, just like the relation which has been seen in this paper.

\bibliography{refs}

\end{document}